\documentclass{llncs}

\usepackage[obeyspaces]{url}
\usepackage{pslatex}
\usepackage{graphicx}
\usepackage{front_cite}

\begin{document}

\title{The Case for Modeling Security, Privacy, Usability and
Reliability (SPUR) in Automotive Software}

\author{K. Venkatesh Prasad \and TJ Giuli \and David Watson}

\institute{Ford Motor Company, Dearborn, MI 48121, USA \\
    \email{\{kprasad,tgiuli,dwatso80\}@ford.com}}

\titlerunning{The Case for Modeling SPUR in Automotive Software}
\idline{Presented at the 2006 Automotive Software Workshop, San Diego, CA\\
        \copyright 2006 Ford Motor Company}

\maketitle

\begin{abstract}
Over the past five years, there has been considerable growth and
established value in the practice of modeling automotive software
requirements.  Much of this growth has been centered on requirements
of software associated with the established functional areas of an
automobile, such as those associated with powertrain, chassis, body,
safety and infotainment. This paper makes a case for modeling four
additional attributes that are increasingly important as vehicles
become information conduits: security,
privacy, usability, and reliability. These four attributes are important in creating
specifications for embedded in-vehicle automotive software.

This paper  examines the automobile context of security, privacy,
usability and reliability (SPUR).  In an automobile safety is
paramount, and while leisure may certainly be found in an
automobile, compromising driver attention could mean loss of life or
serious injury and loss of productivity.  All this makes the
automobile a very special case of information and physical mobility
--- as in the cases of cellular phone based mobility, and mass
transportation (bus, train, ship, airplane) based mobility.

Several real-world use-cases are reviewed to illustrate both the
consumer and system needs associated with SPUR in the automobile and
to highlight the associated functional and non-functional
requirements.  From these requirements the underlying architectural
elements of automotive SPUR are also derived.  Broadly speaking these
elements span three software service domains: the off-board enterprise
software domain, the nomadic (device or service) software domain and
the embedded (in-vehicle) software domain all of which need to work in
tandem for the creation and complete lifecycle management of
automotive software.

\end{abstract}

\section{Introduction}
The nature and terrain of computing in the automobile is in a state
of transition. Automotive computing is transforming from being
function-oriented to being service oriented, while the terrain
(or logical boundaries) of computing {\it in} an automobile is
expanding to include both computing elements in the wireless
external infrastructure and the nomadic (or hand held, mobile)
infrastructure.  This transition is being driven on the one hand by
consumers, wanting to keep pace with their changing life styles and,
on the other hand, by regulatory agencies placing more stringent
demands on the attributes such as safety, emissions, fuel economy.
Given the transformation in the nature and terrain of automotive
computing, this paper makes the case for modeling security, privacy,
usability and reliability (SPUR) --- motivated in part by David
Patterson's manifesto~\cite{SPUR:PattersonDA1}.

For nearly a century, the automobile was defined by components with
local functionality and differentiated by proprietary systems
engineering implementations involving mostly mechanical coupling
between components.  Over the past three decades, with the advent of
microelectronics and local-area
networks~\cite{IntroBib:InVehicleNetCom} in the automobile, there
has been a steady growth in the use of
mechatronics~\cite{Mechatronics2000} and the practice of allocating
functions across multiple components.  The applications of systems
engineering principles, in turn has been extended to heterogenous
contexts of mixed mechanical, electronic, digital, analog (or
discrete-time, continuous-time) sub-systems and components. With the
growing maturity of the software
eco-system~\cite{SoftwareEcosystem:MesserschmittSzyperski} ---
operating systems, programming languages, development environments,
engineering tools, to name some key elements --- the modern automobile
is being increasingly defined by software. There is a trend to
allocate automobile functions across multiple standardized
components (to reduce the number or unique hardware modules) and to
use software design, modeling and engineering for function
implementation and associated product
differentiation~\cite{AUTOSAR}. In this context, the automobile is
rapidly becoming a distributed computing environment.

Commensurate with the growth in demand for new features --- from
both consumers and regulatory agencies --- is the increase in the
complexity of functional allocation across the distributed computing
environment in the vehicle. In addition to the growth in the
complexity of functional allocation, the computing terrain of the
automobile is also rapidly
changing~\cite{NewComputingTerrain:JameelStuempfleJiangFuchs}. With
the advent of wireless personal, local, and wide-area technologies,
the physical boundary of the automobile is no longer the logical
bounding box for functional allocation.  Functions may be
distributed across  on-board computing units~\cite{AUTOSAR},
off-board (such as roadside) infrastructure units~\cite{URL:VII} and
nomadic devices~\cite{Nelson:2004:framework} such as cellular
phones.

To manage this growth in the complexity of allocating functions,
another level of abstraction will likely be required to specify
features enabled by electronics and software. A  service-oriented
computing approach~\cite{Papazouglou:2003:SOC}, is an attractive
option. The present day automobile is function-defined --- most
consumer perceived features are based on the specification of
distributed on-board functions; the future automobile will likely be
service-defined, with features being specified, modeled and
synthesized by aggregating consumer and vehicle related services
from both on-board and off-board sources.

The next section (Section~\ref{sec:SPURInAutoDomain}) of this paper
elaborates the case for SPUR in the automotive context and outlines
the role of modeling SPUR.  Section~\ref{sec:examples} introduces
two broad examples that highlight the new computational terrain of
the automobile and the role of modeling SPUR in these contexts: one
example shows how the computational terrain logically extends from
the the physical boundaries of the automobile into the roadside
infrastructure and the second example illustrates how the new
automotive computational terrain extends through nomadic devices and
services into the wide area communication networks (such as the
wireless telephony networks and, in general, the wireless internet).
Section~\ref{sec:ParkingUseCase} shows how SPUR attributes
associated with a specific use-case could be modeled.
Section~\ref{sec:ModelingRequirements} lists requirements for tools
needed to develop SPUR models.  Section~\ref{sec:Conclusion}, in
conclusion, summarizes the need to model SPUR in the automotive
context.

\section{SPUR in the Automotive Context}
\label{sec:SPURInAutoDomain}

SPUR~\cite{SPUR:PattersonDA1} was advocated on the premise of
shifting research efforts in computer science and engineering away
from making faster, cheaper systems to making systems that are more
secure, privacy-preserving, usable, and reliable.  The automotive
industry is particularly well-suited to understand the value of each
aspect of SPUR-oriented design.

Security in the automotive domain has so far emphasized physical
security. The first automobiles were produced without any built-in
theft deterrents.  Gradually they acquired keys to start the engine
and door locks to protect property left in the vehicle.  Modern
vehicles now use sophisticated radio transmission devices with strong
cryptography to prevent unauthorized entry.

Network connectivity is being added to vehicles through telematics
services (e.g., OnStar,{$\scriptstyle^{\textregistered}$} BMW
ASSIST{$\scriptstyle^{\texttrademark}$}) and hands-free telephony,
introducing the possibility of remote intrusion into a vehicle's
embedded networks. Not only could a remote intrusion compromise the
physical security of the vehicle (i.e., unauthorized remote unlock),
but it could directly affect the vehicle's drivability.  For
example, a virus could trigger the vehicle's theft alarm while
driving. Clearly, as the automotive industry integrates more digital
network technology into vehicles, its impact on both physical and
digital security must be assessed.

On the flip-side of the security coin is a concern for privacy.
Modern vehicles ``know'' much more about their drivers and
passengers than ever before.  Vehicular navigation systems could be
used to correlate data and extract potentially private information.
For example, correlating driver location data with the locations of
points of interest such as stores, places of worship, community
centers and other buildings an organization can build an accurate
profile of the driver's interests.  The privacy concerns of
automobile customers must be treated seriously and safeguarded with
the introduction of new technologies such as telematics and
navigation services.

The usability aspect of SPUR in the automotive context is especially
important because of its impact on safety.  An automobile's
human-machine interface (HMI) must allow the driver to focus on the
task of driving while at the same time providing un-occluded access
to driver information as well as comfort and convenience features
such as climate and radio controls.  Complicating the matter are the
integration of new technologies such as mobile phone services,
voicemail, messaging, and email into the vehicle HMI.  A balance
must be struck between the complexity of an HMI with many features
and safe usability.

Reliability has been a serious concern in the automotive industry
and in the consuming public's minds for some time now.  Automobiles
are increasingly becoming software-driven, not just mechanically
driven.  Therefore, software reliability will be as important as
mechanical reliability in future automobiles.

Table~\ref{tab:AutomotiveSPURExamples} outlines automotive examples
that exhibit varying combinations of SPUR attributes. Each row
categorizes examples as having or lacking some SPUR attributes. For
example, the Carfax{$\scriptstyle^{\textregistered}$} web service
allows anyone to view detailed maintenance and accident histories of
any vehicle for a fee.  The service must be secure to prevent
unauthorized tampering with vehicle records, usable enough for
anyone to understand, and reliable to provide correct information.

\begin{table}
\caption{\label{tab:AutomotiveSPURExamples}Examples illustrating
SPUR in an automotive context and the relative importance (Low,
Medium, High) of each SPUR attribute to each example.}
\begin{center}
\begin{tabular}{lcccc}
Example & S & P & U & R \\ \hline
Carfax{$\scriptstyle{\textregistered}$} database & H & L & H & H \\
Anti-lock braking system & M & L & H & H \\
Door key & H & L & M & H \\
Valet key & H & H & M & L \\
License plate & M & L & L & L \\
\end{tabular}
\end{center}
\end{table}

The examples shown in Table~\ref{tab:AutomotiveSPURExamples} have
software that resides either wholly inside the vehicle, or entirely
outside the vehicle. Conversely, software implementing sophisticated
telematics services reside not only on-board the vehicle but also
off-board, including the IT infrastructure of original equipment
manufacturers (OEMs), dealerships, telecommunications operators, and
in hand held consumer devices. Because of the new push of automotive
software across module and vehicular boundaries, there is a need to
develop models that cross these boundaries as well. Furthermore,
because vehicular telematics software relies on dynamic external
software, models of telematics systems must change along with
deployed systems.  A service-oriented approach to implementing
automotive software --- both in-vehicle software as well as
enterprise software
--- eases the design, implementation and maintenance of systems to ensure
that each requirement of SPUR design is present in the system.

\begin{figure}
  \begin{center}
    \includegraphics[scale=0.4]{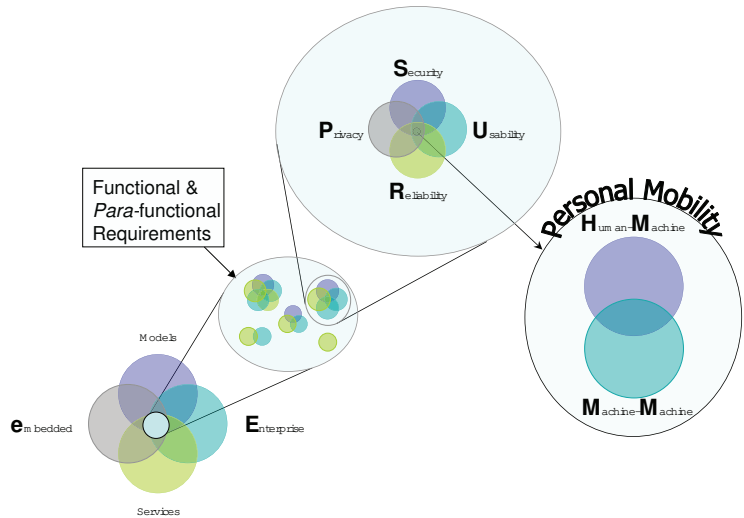}
    \caption{Diagram of automotive SPUR.}
    \label{fig:modeling_space}
  \end{center}
\end{figure}

Figure~\ref{fig:modeling_space} illustrates this interesting space.
As we stated before, we believe it is important to understand how to
model services that cross the embedded and enterprise domains.
Within this space are both functional and para-functional (or
non-functional) requirements.  Functional requirements are more
visible, however we believe that the para-functional requirements
will be increasingly important.  In particular, we are interested in
understanding how the mobility inherent in a vehicle impacts this
space.  Providing functionality to a person driving at highway
speeds requires strong attention to SPUR both at the human to
machine interface as well as the machine to machine interface.  The
safety and quality of the driving experience is clearly affected by
these attributes.  At the same time, designing computer
communications systems that support SPUR concerns in these types of
mobile applications requires careful attention to system
interactions.

%
%

\section{Examples of Automotive Services}
\label{sec:examples}

In this section we use two examples to demonstrate the
trend towards automotive services extending outside the physical
constraints of the vehicle. The first is the Vehicle Infrastructure
Integration (VII) project~\cite{URL:VII}.  The second is the
Vehicle Consumer Services Interface (VCSI)
project~\cite{Nelson:2004:framework}. These two examples demonstrate
integration of the vehicle with roadside infrastructure and consumer
services respectively.

\subsection{VII}

The Vehicle Infrastructure Integration project is a joint effort
involving the United States Department of Transportation (USDOT),
state transportation departments, and vehicle manufactures.  The VII
goal is to develop and deploy the roadside and vehicular
infrastructure needed to improve the safety of the nation's roadways.
By improving the amount and types of information available from the
roadway and by having improved safety warnings and controls, drivers
will be better prepared to mitigate or avoid accidents.  The features
enabled by VII include everything from warning drivers that another
vehicle is about to run a red light, to notifying drivers that a given
section of road is covered with ice. Table~\ref{tab:vii_use_cases}
lists the titles assigned to some of the first scenarios being
considered.  In addition, it highlights how important the SPUR
attributes are to each scenario.  In general, scenarios that are
likely to affect driver behavior or well-being have a high impact from
security.  For example, an incorrect signal that an emergency vehicle
is approaching could cause great headaches to drivers, and potentially
disrupt the usage of this signal by true emergency vehicles.  Thus,
it's important that such a system be secure against malicious
manipulation.  On the other hand, spurious information about traffic
information is less likely to significantly impact drivers, hence it
is listed as having medium importance relative to
security.\footnote{It's important to note that we're talking about a
subjective measure of security for illustrative purposes.  We strongly
believe that all of these attributes are important considerations for
any scenario.}  Privacy is more of a concern when revealing
information about specific vehicles, as in the case of intersection
warnings.  On the other hand, road conditions are likely to be
broadcast to everybody, and therefore unlikely to contain a
significant privacy risk.  In general, useability and reliability are
significant to all of these scenarios.  In some cases, usability is
less important, since the consequences are less severe.

\begin{table}
  \caption{List of VII use cases and the relative importance (Low,
Medium, High) of each SPUR attribute to each use case.}
  \label{tab:vii_use_cases}
  \begin{center}
    \begin{tabular}{lcccc}
      Use case & S & P & U & R \\ \hline
      Emergency Brake Warning & M & L & H & H \\
      Curve Speed Warning & M & L & H & H \\
      Traffic Signal Violation Warning & H & M & H & H \\
      Stop Sign Violation Warning & H & M & H & H \\
      Emergency Vehicle Approaching & H & L & H & H \\
      In-Vehicle Signage & M & L & M & M \\
      Traffic Information and alt route guidance & M & L & M & H \\
      Electronic payments & H & H & M & H \\
      Roadway Condition Information & M & L & H & H \\
      Traffic Management & H & L & H & H \\
      Emergency Vehicle At Scene & H & L & H & H \\
    \end{tabular}
  \end{center}
\end{table}

\subsection{VCSI}

The second project, the Vehicle Consumer Services Interface (VCSI), is
a project at Ford to provide an interface between consumers, their
personal devices, off-board services, and vehicle systems including
both networks and devices.  To demonstrate this system, we developed a
prototype vehicle that contained several specific applications
including those shown in Figure~\ref{tab:vcsi_services}.  As with the
VII examples above, we've made some attempt to demonstrate the
relative importance of each SPUR attribute to each service.  Since
most of the consumer facing services provided by VCSI are not safety
critical, they have lower requirements on usability and reliability.
At the same time, most of these services depend on interfacing with
devices that have personal information.  In that context, it's
important that the privacy of the data contained within those devices
be kept secure.

\begin{table}
  \caption{List of VCSI services and the relative importance (Low,
Medium, High) of each SPUR attribute to each service.}
  \label{tab:vcsi_services}
  \begin{center}
    \begin{tabular}{lcccc}
      Service & S & P & U & R \\ \hline
      Vehicle Personalization & L & H & M & H \\
      Personal Information Management & H & H & M & M \\
      MyHome (Home Automation Services) & H & H & M & M \\
      Bluetooth Technology & H & H & M & M \\
      Real-time navigation & M & L & M & H \\
      Diagnostics & H & M & H & H \\
      In-vehicle media player & M & M & M & M \\
    \end{tabular}
  \end{center}
\end{table}

Overall, we think these two projects demonstrate an increasing trend
towards increased connectivity with a vehicle, both from consumer
devices and from roadside infrastructure.  In addition, we believe
that modeling provides the means to understand these services provided
to the consumer at a system level.

\section{Electronic Payment Use Case}
\label{sec:ParkingUseCase}

\begin{figure}
  \begin{center}
    \includegraphics[width=1.0\columnwidth,clip=true,viewport= .0in 2in 8in 11in]{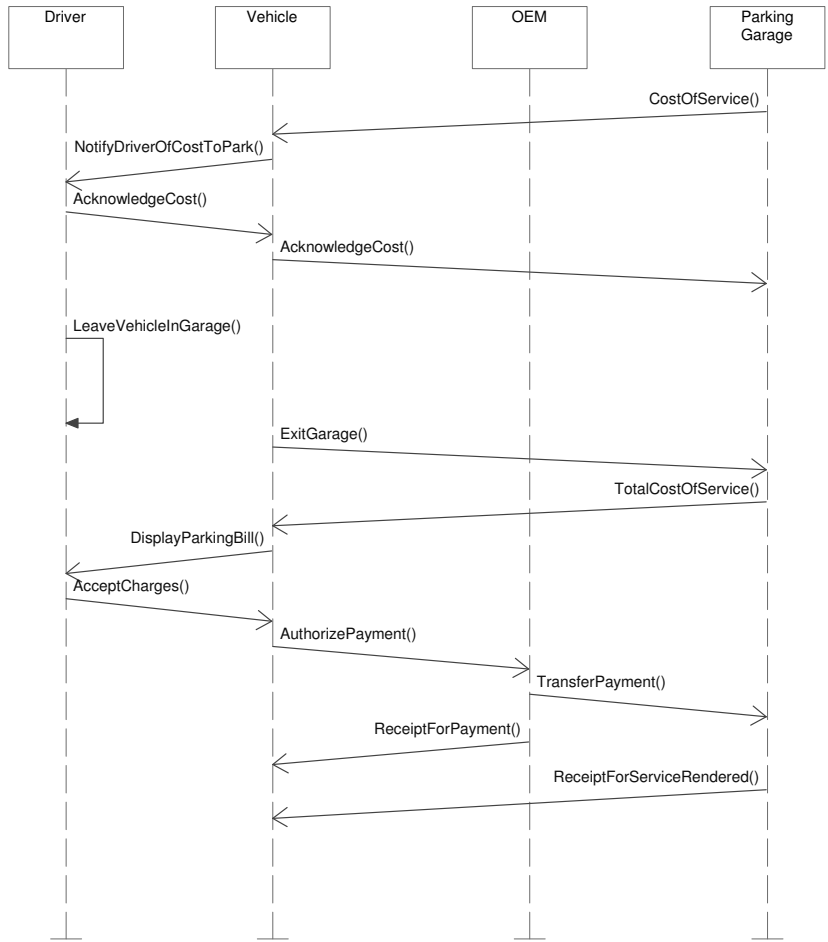}
    \caption{A sequence diagram showing the interactions between entities in a
    parking garage with an electronic payment service.  In this scenario a driver
    parks her car in a smart parking garage and electronically pays upon exit.}
    \label{fig:ParkingUseCase}
  \end{center}
\end{figure}

In this section, we present a more in-depth look at the electronic
payment use case mentioned in Section~\ref{sec:examples} and how it
relates to SPUR-oriented design. With electronic payments, drivers
will have the ability to pay for parking electronically without
interacting with a parking meter or a garage attendant. Drivers will
no longer have to dig around for spare change and municipalities
will no longer have to collect cash from parking meters.

\begin{figure}
\centering
    \rotatebox{270}{\includegraphics[scale=0.3]{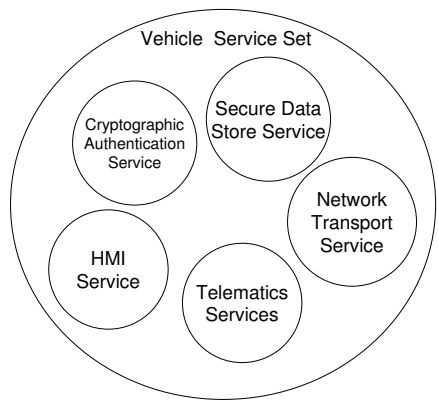}}
    \caption{The vehicle services needed to implement electronic
    payment in a service-oriented architecture.}
    \label{fig:VehicleServices}
\end{figure}

Figure~\ref{fig:ParkingUseCase} shows a sequence diagram for a
vehicle involved in an electronic payment scenario with a parking
garage.  The main entities in the diagram are the driver of the
vehicle, the vehicle's software systems (implemented in a
service-oriented architecture, as shown in
Figure~\ref{fig:VehicleServices}), the vehicle's OEM (or a delegate
of the OEM), and the parking garage authority. When the vehicle
enters the garage, the garage transmits a list of services and their
costs to the vehicle, which in turn presents this information to the
driver through the vehicle's HMI.

Assuming that the driver is willing to pay the cost to park, she
acknowledges the cost of service, parks the vehicle and leaves.  The
vehicle sends a signed acknowledgement to the garage. Later, the
driver returns and begins driving out of the garage. The garage
calculates the amount of money owed and securely transmits a bill to
the vehicle. The vehicle notifies the driver of how much is owed
through the HMI and requests that the driver consent to pay.
Confirmation from the driver causes the vehicle to transmit an
encrypted, signed payment authorization message to the OEM.  The
OEM, acting in the role of an e-payment service, securely credits
the funds to the parking garage and returns a signed receipt to the
vehicle showing proof of payment. Finally, the garage sends a signed
receipt to the vehicle showing that it has received its requested
payment.

Thus, at the end of the interaction between the driver and the
garage, the driver has proof from both the OEM and the garage that
she has paid what she owed.  The garage has a signed acknowledgement
from the driver stating that she understood the cost to park before
she parked her vehicle as well as funds deposited by the OEM to pay
for parking.  The receipts returned to the driver are necessary to
prove that she paid for services in the case of a dispute between
the garage and driver.  Similarly, the signed acknowledgement
agreeing to the cost of parking from the driver is necessary to
dissuade a driver from reneging on payment upon exit from the
garage.

\subsection{Challenges}

There are many challenges involving SPUR in the context of such an
automotive e-payment system.  While many of these challenges are not
unique to e-payments in general, the scope of this paper is to
understand how these issues are unique in an automotive context.

First are questions of infrastructure.  E-payments require a secure,
potentially private, system for transferring money from a driver or
other occupant in the car to a specific payee.  We also assume that
these payments will reflect current cash payment characteristics,
specifically, we need to support individual transactions of less
than one dollar.  This requires the support of a third party to
aggregate payments on both sides of the payment.  This could be the
vehicle manufacture, as we've outlined before, a credit card issuer,
or an Internet e-payment provider.

Automotive e-payment is inherently a mobile application.  Malicious
agents are likely to have easy access to all communication that
takes place outside the vehicle.  In addition, unlike personal
mobile devices such as a cell phone, there is inherently less
physical security over the vehicle.  Cars are often parked in public
spaces, and routinely in control of mechanics.  Even users sometime
have a vested interest in modifying the vehicle software, as
evidenced by powertrain modification chips.  These reasons imply
that some type of end-to-end assurance is needed about the
legitimacy of each individual transaction.  However, there is an
inherent trade-off between the sophistication of a given security
system and the risk of compromise. For example, an individual driver
is unlikely to notice or care if a individual penny or quarter is
missing from his car when she takes it in for service.  Similarly,
users often trade off convenience for increased risk of monetary
loss.  For example, many electronic cash cards such as the Octopus
card used in Hong Kong~\cite{PaynterOctopus} require no
authentication to use, and the owner assumes that a lost card
implies the money associated with that card is also lost. Similarly,
in-vehicle e-payment systems need to take into account the unique
environment when trading off risk with cost. Mobility also has
implications for the reliability of the system. There is no
guarantee that a device will always stay in communications range
during the period of a transaction.

Providing security and privacy in electronic transactions naturally
implies the use of cryptographic protocols.  In contrast to general
purpose computers, the computational power and upgrade capabilities
of embedded devices is severely limited.  In addition, unlike the
consumer electronics side of the embedded, mobile marketplace,
vehicle software has a useful life of over ten years.  In this
context, how do we ensure that the computational power will be great
enough to support key lengths that can't be easily compromised long
into the future, without needless expense?  At the same time, flaws
in cryptographic protocols are not uncommon, so the in-vehicle
software should be upgradable, without causing undue burden on the
driver.

Second, are questions of authentication.  How do we authenticate
that the person responsible for the account used in the transaction
is authorized to make the payment?  We can't always assume that the
driver is authorized to make payments with an account associated
with the vehicle.  Valets or even teenage drivers quickly complicate
this assumption.  At the same time, we want to authenticate the
payee to the driver, making sure that a hacker hasn't set up their
own virtual toll booth at the side of the highway, while still
making it easy for small businesses to use the system.  In some
sense, the physical nature of our scenario provides opportunities
not usually seen on the Internet.  Most drivers require a physical
or electronic key in order to enter a vehicle.  At the same time, in
the scenarios that we described, the payee will be in physical view
of the driver.  This presents an opportunity to provide out-of-band
signaling to facilitate authentication.

Similarly, the physical nature of owning a vehicle presents an
opportunity for associating real people with digital identities.  In
buying or leasing a vehicle, most buyers have little expectation of
privacy.  Most transactions require some type of financing,
necessitating at least a credit check.  Even in situations where this
isn't the case (e.g. person to person cash transactions), owning a
vehicle requires licensing with the state, another transaction which
implies a lack of privacy, and a financial interest in correctly
identifying the owner.

Finally, the interface between the driver and the vehicle computer
system poses several important challenges.  Because we are talking
about the driver authorizing payments while driving, this interaction
needs to require little attention from the driver.  At the same time,
we need drivers to understand the security implications of the actions
they're performing.  Studies of web browser security have demonstrated
techniques to better inform users of the security implications of the
current browser state~\cite{Ye:2005:trusted_paths}.

\section{Modeling Requirements}
\label{sec:ModelingRequirements}

The electronic payment use case detailed in
Section~\ref{sec:ParkingUseCase} touches on all aspects of
SPUR-oriented design.  For vehicular  electronic payment to be
widely accepted, sensitive financial information must be securely
exchanged between the vehicle, the OEM, and a service vendor.  The
privacy of financial dealings must also be preserved.  Furthermore,
the HMI must clearly present information about the cost of a service
and indicate when consent is required.  Finally, electronic payment
systems must be reliable enough to give drivers the confidence to
wholly adopt them.

Modeling the parking garage use case requires a diverse set of tools
and disciplines.  The driver must not be distracted while making
financial transactions yet the HMI must be involving enough to
assure the driver that they are making a secure transaction.  The
HMI may use a text display, an LCD, voice recognition, or a
combination of interface technologies to communicate with the
driver.  We must be able to realistically model a user interface with
all of these qualities.

A significant amount of software of varying complexity is involved
in our use case, from less complex programs embedded in the vehicle
to highly complex back-end software at the OEM and parking garage
vendor. The interactions between the vehicle, the OEM, and the
service vendor must be modeled as well.  We thus require a software
modeling tool that can effectively model heterogeneous software
environments with varying levels of complexity.

Each aspect of SPUR is a whole-system attribute.  For example,
spending resources on creating a security-hardened implementation of
the vehicle's embedded programs is useless if the communications
between the vehicle and OEM are unencrypted.  Similarly, an electronic
payment system with a highly reliable embedded program but a buggy OEM
back-end interface makes the system as a whole unreliable.

Therefore, to fully evaluate each aspect of SPUR we must be able to
study the HMI of the vehicle, its embedded programs, the OEM and
parking garage enterprise software as a single system.  We require a
single tool or suite of tools that can fully inter-operate in order
to model the interactions between each of the system's components.
The tool must allow us to inject faults or directed attacks and
measure the effects both in terms of software metrics (i.e. loss of
privacy, reduced reliability) and in terms of customer-facing
metrics such as the effect of a fault at the OEM on the in-vehicle
HMI.

\section{Conclusion}
\label{sec:Conclusion}

Given the transformation that both the nature and terrain of
computing in the automobile are undergoing, this paper has outlined
the case to model security, privacy, usability and reliability
(SPUR) in the context of the software enabled services associated
with the automobile.  SPUR represents a set of attributes that are
not explicitly articulated or demanded by the end customer or
consumer and hence, broadly speaking, SPUR represents
non-functional, or para-functional, attributes.

Security, privacy, usability and reliability have all been product
creation requirements that have been well understood and refined by
the automotive industry over the years, but almost exclusively in
the mechanical or physical context. With the advent of the
information-enabled automobile --- connected to the roadside
infrastructure and to consumer devices --- SPUR takes on a very
different interpretation.  This paper highlights the importance of
SPUR.  In addition, we make a case for modeling SPUR, as this would
avoid costly and time consuming hardware investments and  will
likely provide quick insights into how  technologies and standards
could be adapted to meet automotive SPUR requirements.

\bibliographystyle{splncs}

\bibliography{SPURBiB}

\end{document}